\documentclass[conference]{IEEEtran}
\IEEEoverridecommandlockouts
\usepackage{cite}
\usepackage[belowskip=-10pt,aboveskip=0pt,small]{caption}
\IEEEaftertitletext{\vspace{-1.5\baselineskip}}
\usepackage{amsmath,amssymb,amsfonts,nccmath,placeins}
\usepackage{algorithmic}
\usepackage{graphicx}
\usepackage{textcomp}
\usepackage{xcolor}
\def\BibTeX{{\rm B\kern-.05em{\sc i\kern-.025em b}\kern-.08em
    T\kern-.1667em\lower.7ex\hbox{E}\kern-.125emX}}
\begin{document}

\title{Polar Coded Faster-than-Nyquist (FTN) Signaling with Symbol-by-Symbol Detection\\
}
\author{
\IEEEauthorblockN{Abdulsamet Caglan}
\IEEEauthorblockA{\textit{Electronic Engineering Dept.} \\
\textit{Gebze Technical University}\\
Gebze, Turkey \\
abdulsamet.caglan@tubitak.gov.tr}
\and
\IEEEauthorblockN{Adem Cicek}
\IEEEauthorblockA{\textit{Electrical and Electronic Engineering Dept.} \\
\textit{Ankara Yildirim Beyazit University}\\
Ankara, Turkey \\
acicek@ybu.edu.tr}
\and
\IEEEauthorblockN{Enver Cavus}
\IEEEauthorblockA{\textit{Electrical and Electronic Engineering Dept.} \\
\textit{Ankara Yildirim Beyazit University}\\
Ankara, Turkey \\
ecavus@ybu.edu.tr}
\and
\IEEEauthorblockN{\hspace*{2.0cm}Ebrahim Bedeer}
\IEEEauthorblockA{\textit{\hspace*{2.0cm}Electrical and Computer Engineering Dept.} \\
\hspace*{2.0cm}\textit{University of Saskatchewan}\\
\hspace*{2.0cm}Saskatoon, SK, Canada \\
\hspace*{2.0cm}e.bedeer@usask.ca}
\and
\IEEEauthorblockN{\hspace*{-2cm}Halim Yanikomeroglu}
\IEEEauthorblockA{\hspace*{-2cm}\textit{System and Computer Engineering Dept.} \\
\hspace*{-2cm}\textit{Carleton University}\\
\hspace*{-2cm}Ottawa, ON, Canada \\
\hspace*{-2cm}halim.yanikomeroglu@sce.carleton.ca}
}
\maketitle
\begin{abstract}
 Reduced complexity faster-than-Nyquist (FTN) signaling systems are gaining increased attention as they  provide improved bandwidth utilization for an acceptable level of detection complexity. In order to have a better understanding of the tradeoff between performance and complexity of the reduced complexity FTN detection techniques, it is necessary to study these techniques in the presence of channel coding. In this paper, we investigate the performance a polar coded FTN system which uses a reduced complexity FTN detection, namely, the recently proposed ``successive symbol-by-symbol with go-back-$K$ sequence estimation (SSSgb$K$SE)'' technique. Simulations are performed for various intersymbol-interference (ISI) levels and for various go-back-$K$ values. Bit error rate (BER) performance of Bahl-Cocke-Jelinek-Raviv (BCJR) detection and SSSgb$K$SE detection techniques are studied for both uncoded and polar coded systems. Simulation results reveal that polar codes can compensate some of the performance loss incurred in the reduced complexity SSSgb$K$SE technique and assist in closing the performance gap between BCJR and SSSgb$K$SE detection algorithms.
\end{abstract}

\section{Introduction}
Dramatic growth in demand for data services necessitates significant improvement in the bandwidth utilization. One promising technique to accommodate the increasing demand for spectrum efficiency is faster-than-Nyquist (FTN) signaling. The traditional Nyquist rate transmission adopts an orthogonal signaling scheme which avoids intersymbol-interference (ISI) to simplify the design of the receiver.
FTN signaling relaxes the orthogonality constraint by reducing the time interval between successive transmit pulses. Consequently, symbols are transmitted in a higher rate than the Nyquist limit, and hence, a controlled ISI is introduced at the receiver.
Mazo introduced FTN signaling in \cite{Mazo}, where he showed that the minimum distance of uncoded binary sinc pulses does not change when they are packed up to 0.802 of the symbol duration. This constraint on the minimum distance is known as Mazo limit \cite{Mazo} which basically reveals the possibility of transmitting 25\% more bits in the same bandwidth and energy without any performance loss.
\par
While the non-orthogonality of FTN signaling provides bandwidth efficiency, it also introduces ISI at the receiver. The non-zero ISI complicates the detection process and requires sequence estimation methods to recover the transmitted symbols. In \cite{Liveris}, the Viterbi algorithm (VA) that is based on maximum likelihood sequence estimation (MLSE) is used to detect FTN signaling. To reduce complexity, truncated VA and reduced states Bahl-Cocke-Jelinek-Raviv (BCJR) algorithms are proposed in \cite{Anderson4} and \cite{Anderson2}, respectively. Although these algorithms are very effective even in severe ISI scenarios, they have a high computational complexity which exponentially increases with ISI length. To further reduce the computational complexities of FTN detection algorithms, different studies are reported in the literature (see e.g., \cite{Anderson3} -- \cite{Bedeer4} and references therein). Sequence estimation methods for high-order quadrature amplitude modulations (QAM) and phase shift keying (PSK) are reported with polynomial time complexity in \cite{Bedeer3} and \cite{Bedeer4}, respectively. In \cite{Bedeer} a successive symbol-by-symbol with go-back-$K$ sequence estimation (SSSgb$K$SE) algorithm is proposed. 
In \cite{Kulhandjian}, two probabilistic data association (PDA) algorithms based on Gaussian separability method are introduced to detect binary phase-shift keying (BPSK) FTN signaling in polynomial time complexity. 
\par
In general, most of the reduced complexity FTN signaling studies in literature focused on detection algorithms alone and their interaction with channel codes has been overlooked. When a channel code is included in the system either no reduced complexity FTN detection algorithm is considered \cite{LDPC}, \cite{Polar_Turbo} or BCJR detection algorithm is employed \cite{Turbo} -- \cite{MMSE}. 
\begin{figure*}[t]
\centering
\includegraphics[width=11cm,height=5cm]{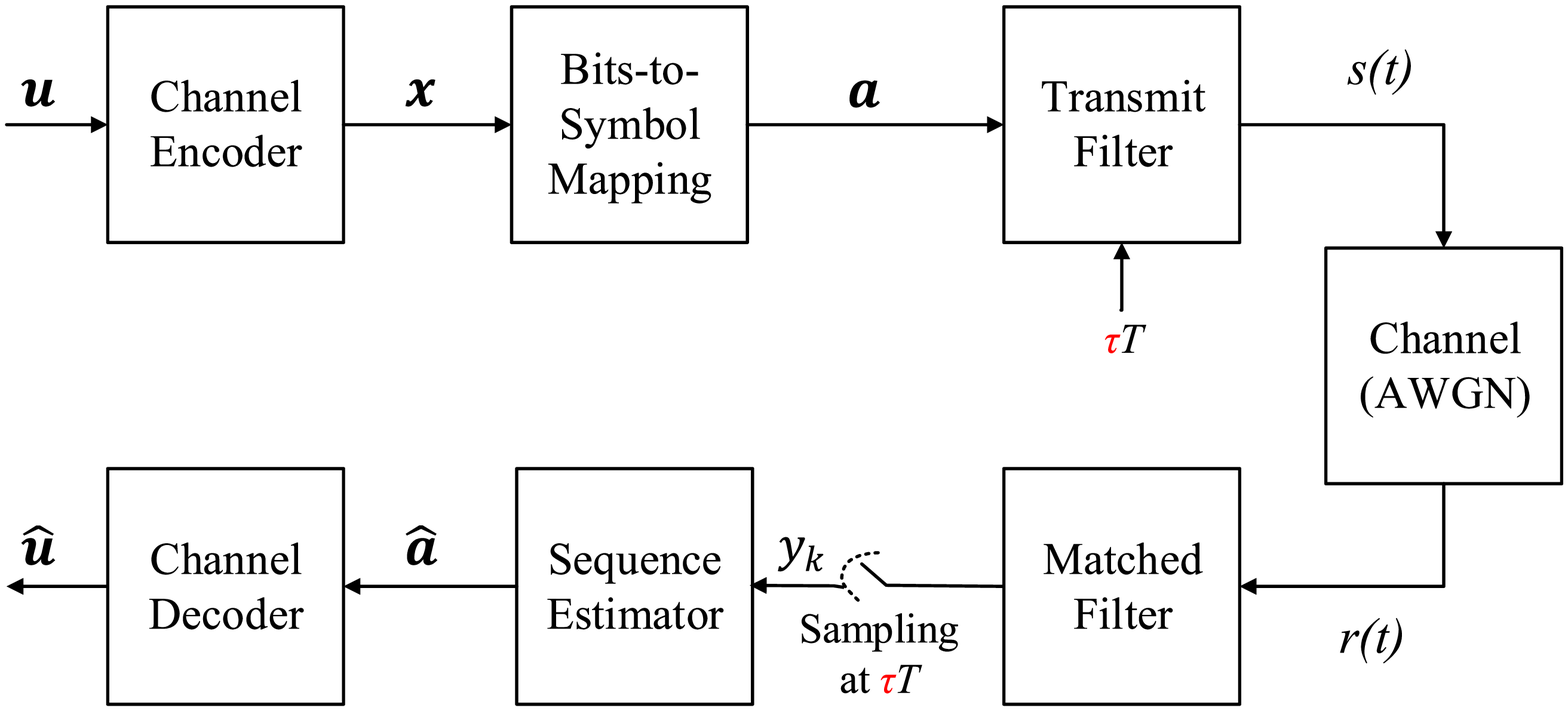}
\caption{Block diagram of FTN signaling with channel coding. \label{A}}
\end{figure*}

In this work, we focus on the performance of the reduced complexity FTN detection algorithms in a channel coded communication system. Specifically, for various go-back-$K$ values and different ISI levels, bit error rate (BER) performance of SSSgb$K$SE detection algorithm is studied with polar codes and its simulation results are compared with BCJR algorithm, which is considered as a benchmark in this paper. The simulation results show that polar codes can compensate for some of the performance loss incurred in the reduced complexity SSSgb$K$SE technique. Additionally, the results reveal that under light ISI conditions, polar codes considerably improve the performance of the SSSgb$K$SE algorithm and bridges the performance gap with the BCJR algorithm. 
\par
The remainder of this paper is organized as follows. Section II presents the system model of coded FTN signaling system. In Section III, the proposed polar coded FTN system with SSSgb$K$SE is introduced and details of SSSgb$K$SE algorithm and polar code decoding are given. Simulation results are presented in Section IV. Finally, the paper is concluded in Section V.
\section{FTN Signaling System Model}
\par
A coded communication system employing FTN signaling is considered in this work. The block diagram of the system is illustrated in Fig. \ref{A}. At the transmitter side, $M$-bit information vector $\boldsymbol{u}$ is first encoded into code-words \textit{\textbf{x}} of length $N$ using a channel encoder, where $M/N$ denotes the code rate. Then the encoded bits are converted to binary phase shift keying (BPSK) or quadrature phase shift keying (QPSK) modulated symbols ${a}_n$ in bits-to-symbols mapping block where $0<n\leq N$. Finally, the modulated symbols are shaped by a unit-energy pulse $p(t)$ and then passed through to a transmit filter block where FTN signaling is applied, i.e., transmitting a pulse carrying a data symbol every $\tau T$, where $0<\tau\leq1$ is the time packing/acceleration parameter and $T$ is the symbol duration. The transmitted signal $s(t)$ can be expressed as
\begin{equation}
s(t) = \sqrt{E_s}\sum_{n=1}^{N}a_n p(t-n\tau T), \quad 0\:<\tau\leq\:1,
\end{equation}
where $E_s$ is the symbol energy. 
In the channel, $s(t)$ is exposed to additive white Gaussian noise (AWGN), $n(t)$, with zero mean and variance $\sigma^2$. The received FTN signal is written as
\begin{equation}
r(t)\:=\:s(t)\:+\:n(t).
\end{equation}
The received FTN signal, $r(t)$, is passed through a matched filter block at the receiver as shown in Fig. \ref{A}. The received signal after the matched filter can be written as
\begin{equation}
y(t) = \sqrt{E_s}\sum_{n=1}^{N}a_n g(t-n\tau T)+w(t),
\end{equation}
where $g(t)=\int p(x)p(x-t)dx$ and $w(t) = \int n(x) p(x-t)dx$. 

\par
The output signal of the matched filter is passed through a sampler. As the orthogonality is related to the pulses carrying the data symbols are not preserved, ISI occurs between the received samples. The output signal of the matched-filter is sampled every $\tau T$ and can be expressed as 
\begin{equation}
\begin{array}{l}
y_k=y(k\tau T) \\
=\sqrt{E_s}\sum_{n=1}^{N}a_n g(k\tau T - n\tau T) + w(k\tau T)\\
=\underbrace{\sqrt{E_s} a_k g(0)}_\text{desired symbol}
+\underbrace{\sqrt{E_s}\sum_{n=1,n\neq k}^{N} a_n g((k-n)\tau T)}_\text{ISI from adjacent symbols}\\
\ \ +w(k\tau T).
\end{array}
\end{equation}
Accordingly, sequence estimation techniques are needed to remove the ISI and to estimate the transmitted data symbols. In sequence estimation, estimated symbol values $\hat{a}_n$ are calculated and passed through the channel decoder block. In channel decoder block, log-likelihood ratio (LLR) values are calculated and data bits are evaluated based on soft decoding technique.
\section{Polar Coded Reduced Complexity FTN Detection}
\par
In this section, we study a polar coded FTN system with SSSgb$K$SE detection. At the transmitter, information bits are passed through a polar encoder, which is constructed based on the channel polarization concept \cite{Arikan}. Channel polarization is achieved by recursively applying linear transformations to initially independent binary erasure channels (BEC) \cite{Sural}, $W$,  which finally transforms the channel into reliable and unreliable bit-channels. In the following, this process is briefly summarized for the erasure channels. Channel polarization process starts with initializing channel erasure probability, $\epsilon=0.5$, and symmetric capacity as
\begin{equation}
I(W^{i\:=\:1}_{N\:=\:1}) = 1-\epsilon = 0.5,
\end{equation}
where $W$ denotes a set of channels and $i$ is the bit index bounded with $1\leq i\leq N$. Then, $N$ channel symmetric capacities are calculated with recursive formulas as shown in (6) and (7),
\begin{equation}
I(W_N^{(2i-1)}) = I(W_{N/2}^{(i)})^2, \end{equation}
\begin{equation}
I(W_N^{(2i)}) = 2I(W_{N/2}^{(i)})-I(W_{N/2}^{(i)})^2.
\end{equation}
After $N$ channel symmetric capacities are calculated, the highest $M$ capacities over $N$ channels are selected as free bits indices, which are considered as the most reliable channels and information bits are inserted into these indices. The remaining channel positions are denoted as frozen indices, which are mapped to zero. For example, let $\epsilon=0.5$, $N=4$ and $M=2$. Message is two bit $``11"$.
\newline
\newline
\hspace*{2.5cm}$I(W_4^{1}) = 0.0625,$\: \text{(frozen)},\\ 
\hspace*{2.5cm}$I(W_4^{2}) = 0.4375,$\: \text{(frozen)},\\
\hspace*{2.5cm}$I(W_4^{3}) = 0.5625,$\: \text{(free)},\\
\hspace*{2.5cm}$I(W_4^{4}) = 0.9375,$\: \text{(free)}.\\
\newline
As can be seen from channel symmetric capacity values, the two highest capacities are 3 and 4. So information bits are inserted into these bit indices, and the 1 and 2 indices are frozen bits and mapped to zero, such that uncoded word $d=[0011]$ is created. Complexity of channel polarization process is presented as \textit{O(N)}. In the encoder block, $d$ is multiplied with a generator matrix as shown in (9). Generator matrix $H$ is created by
$ \rm{log}_{2} \textit{N} $ to the Kronecker-power of $F=
\begin{pmatrix} 
1 & 0\\ 
1 & 1 
\end{pmatrix}$ matrix as
\begin{equation}
H= F^{\rm{kron}(\rm{log}_{2} \textit{N})},
\end{equation}
\begin{equation}
\textit{\textbf{x}} = dH.
\end{equation}
After generating codeword vector \textit{\textbf{x}}, codeword bits are passed through the bits-to-symbol mapping block for BPSK/QPSK modulation and transmit filter block for FTN signaling. At the receiver, after the matched filter and sampling, the received symbols are passed to the sequence estimation block. 
\par
In SSSgbKSE algorithm\cite{Bedeer}, the received $k$th sample value $y_k$ is expressed as
\begin{equation}
\begin{array}{l}
y_k=\underbrace{G_{1,L}a_{k-L+1}+...+\underbrace{G_{1,K+1}a_{k-K}+...+G_{1,2}a_{k-1}}_\text{Previous $K$ symbols to be re-estimated}}_\text{ISI from previous $L-1$ symbols} \\
\qquad\qquad\qquad+\underbrace{G_{1,1}a_k}_\text{Current symbol to be estimated}\\
\qquad\qquad\qquad+\underbrace{G_{1,2}a_{k+1}+...+G_{1,L}a_{k+L-1},}_\text{ISI from upcoming $L-1$ symbols} \\
\end{array}
\end{equation}
where $G$ is the ISI matrix and is defined in \cite{Bedeer1}. Hence, the improved re-estimation of the $(k - K)$th data symbol can be written as
\begin{figure}[t]
\centering
\includegraphics[width=88mm,height=60mm]{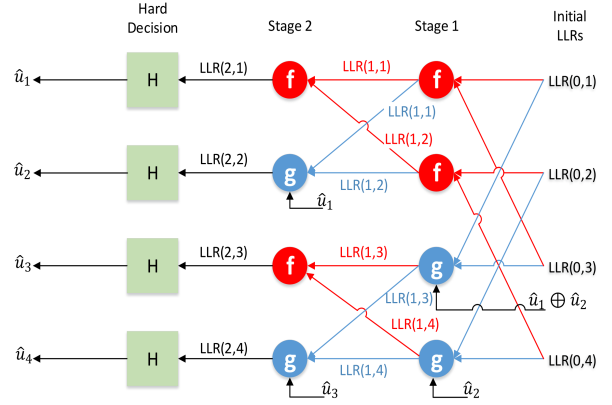}
\caption{Successive cancellation decoder of polar codes.\label{SCD}}
\end{figure}
\begin{equation}
\begin{array}{l}
\hat{a}_{k-K}=y_{k-K} \\
\ \ \ \ \ -\underbrace{(G_{1,L}\hat{a}_{k-K-L+1}+...+G_{1,2}\hat{a}_{k-K-1})}_\text{ISI from previous $L-1$ data symbols of the $(k-K)$th data symbols} \\
\ \ \ \ \ -\underbrace{(G_{1,2}\hat{\hat{a}}_{k+1}+...+G_{1,K+1}\hat{\hat{a}}_{k}).}_\text{ISI from upcoming $K$ data symbols of the $(k-K)$ the data symbol} \\
\end{array}
\end{equation}

\par
After employing SSSgb$K$SE algorithm, the soft symbol values obtained without quantization  $\hat{a}_k$ is given as
\begin{equation}
\begin{array}{l}
\hat{a}_k=(y_k\\
-\underbrace{(G_{1,L}\hat{a}_{k-L+1}+...
+\underbrace{G_{1,K+1}\hat{\hat{a}}_{k-K}+...+G_{1,2}\hat{\hat{a}}_{k-1})).}_{\substack{\text{ISI from previous $K$ data symbols} \\ \text{with improved estimation accuracy}}}}_\text{ISI from previous $L-1$ data symbols}
\end{array}
\end{equation}
Soft symbols $\hat{a}_k$ estimated using SSSgb$K$SE are then inputted as LLR values to the polar decoder. In polar decoder, successive cancellation decoding (SCD) algorithm \cite{Yuan} is used. SCD algorithm decodes the data bits at $ s=\log_{2} N $ stages, where $N$ is the codeword size of the used polar code. A simple example of SCD algorithm is illustrated in Fig. \ref{SCD} for $N=4$ ($s=2$ stages).
In Fig. \ref{SCD}, two different operations are employed at SCD. These operations are denoted as $f$ and $g$ nodes in Fig. \ref{SCD}. Eqn. (13) is employed at the node $f$ while Eqn. (14) is used at the node $g$ where $u_{\rm{sum}}$ is partial sum of previous decoded bits. At every stage, log-likelihood ratios are updated as
\begin{align*}
m={\rm{LLR}}(i,j), \;
n={\rm{LLR}}(i,j+N/2^{i+1}),
\end{align*}
\begin{equation}
{\rm{LLR}}(i+1,j) = {\rm{sign}}(m){\rm{sign}}(n){\rm{min}}(|m|,|n|),
\end{equation}
\begin{align*}
p={\rm{LLR}}(i,j-N/2^{i+1}), \;
r={\rm{LLR}}(i,j),
\end{align*}
\begin{equation}
{\rm{LLR}}(i+1,j) = p(-1)^{u_{\rm{sum}}}+r.
\end{equation}
\begin{figure}[t]
\centering
\includegraphics[width=88mm,height=70mm]{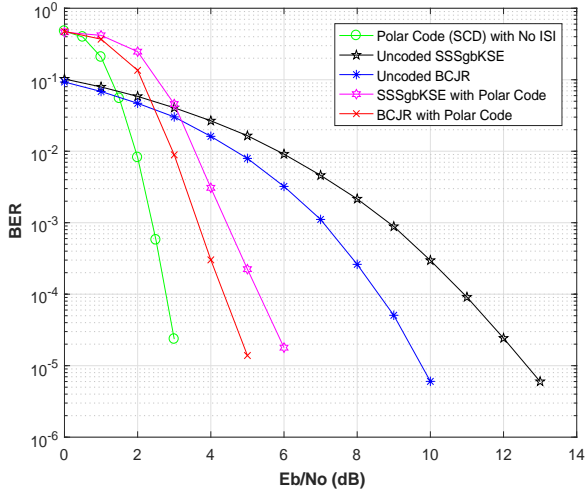}
\caption{BER performance of uncoded and coded FTN detection systems for $\tau=$\:0.8. \label{D}}
\end{figure}
\par
Finally decoded bits, $\hat{u}$, are determined using hard decision operation as
\begin{equation}
\hat{u}_i = \begin{cases}
0\:, & \text{if frozen bit}, \\
0\:, & \text{if free bit and LLR(i)} \geq 0,\\  
1\:, & \text{if free bit and LLR(i)}<0.
\end{cases}
\end{equation}
\par
In SCD, as the node $g$ operations need previously decoded bits, decoded operation is carried out for one bit at a time. 
\section{Simulation Results}
\par
The polar coded FTN system with SSSgb$K$SE detection algorithm is simulated in Matlab and its performance is compared with BCJR algorithm. BCJR algorithm is selected for comparison as it approximates the optimal detection of FTN signaling. BER performance of the system is studied using BPSK modulation in an AWGN channel for the values of $\mathbf{\tau =}$ 0.8, 0.7 and 0.6. In simulations, a polar code with code length $N$=1024 and code rate 1/2 is employed with a SCD algorithm of polar code. In the transmit filter, a root raised cosine filter with a roll-off factor of $\beta = 0.3$ is assumed. In SSSgb$K$SE algorithm, a go-back parameter of $K=1$ is used unless otherwise mentioned. Also, the spectral efficiency (SE) is calculated using (\ref{se_formula}) for channel coded baseband BPSK. 
\begin{equation}\label{se_formula}
\textup{SE} = \frac{\frac{M}{N}\frac{1}{\tau T}}{{\frac{1}{2T}(1+\beta)}}.
\end{equation}
The SE for the cases of $\tau = 0.6, 0.7$ and $0.8$ are 1.28, 1.1 and 0.96 bit/s/Hz respectively, while it is 0.77 bit/s/Hz for $\tau = 1$ (i.e. no ISI). Note that polar encoder and decoder has complexity of $O(Nlog(N))$. The FTN detection algorithm SSSgb$K$SE has $O(L)$ complexity, whereas BCJR detection algorithm has an exponential complexity with ISI length.
\par
Fig. \ref{D} compares the BER performance of BCJR and SSSgb$K$SE detection algorithms for the polar coded and uncoded cases when $\mathbf{\tau =}$ 0.8. In Fig \ref{D}, the BER performance for a polar code with no ISI is also included for the sake of completeness. Note that in the following all performance comparisons between different curves are stated with respect to the BER of $10^{-4}$. For the uncoded case, compared to the SSSgb$K$SE detection algorithm the BCJR algorithm provides $2.5\:$dB performance gain. When a polar code is employed, the performance loss between SSSgb$K$SE and BCJR algorithms is reduced to $1\:$dB at BER of $10^{-4}$. Clearly, in a coded system, polar code compensates the performance loss introduced by
\begin{figure}[t]
\centering
\includegraphics[width=88mm,height=70mm]{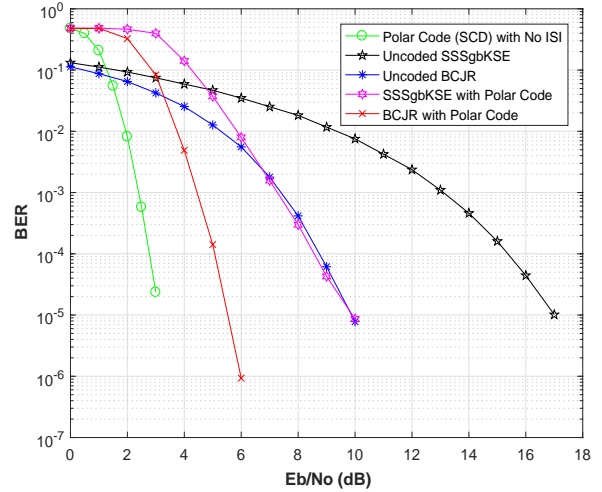}
\caption{BER performance of uncoded and coded FTN detection systems for $\tau=$\:0.7. \label{C}}
\end{figure}
\begin{figure}[t]
\centering
\includegraphics[width=88mm,height=70mm]{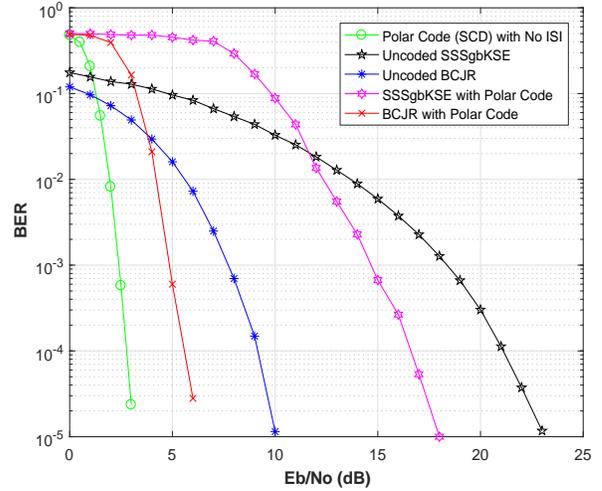}
\caption{BER performance of uncoded and coded FTN detection systems for $\tau=$\:0.6. \label{B}}
\end{figure}
the SSSgb$K$SE detection algorithm , and it is able to provide an additional $1.5$\:dB performance gain to SSSgb$K$SE detection system compared to coded BCJR detection system.
\par
Figs. \ref{C} and \ref{B} illustrate the BER performances when $\tau=$\:0.7 and $\tau=$\:0.6, respectively. As expected, when ISI gets stronger with decreasing the $\tau$ value, the performance gap between uncoded BCJR detection and uncoded SSSgb$K$SE detection systems increases. In Fig. \ref{C}, 
performance loss of uncoded SSSgb$K$SE detection system is $6.6$\:dB while the performance gap between uncoded systems extends to $12$\:dB in Fig. \ref{B}. When polar codes is integrated to the FTN system, the performance gain of BCJR detection system with polar codes compared to the SSSgb$K$SE detection system with polar code is reduced to $3.5$\:dB (from $6.6$\:dB) for $\tau=$\:0.7 and  $11$\:dB (from $12$\:dB) for $\tau=$\:0.6. 
\begin{figure}[ht]
\centering
\includegraphics[width=88mm,height=70mm]{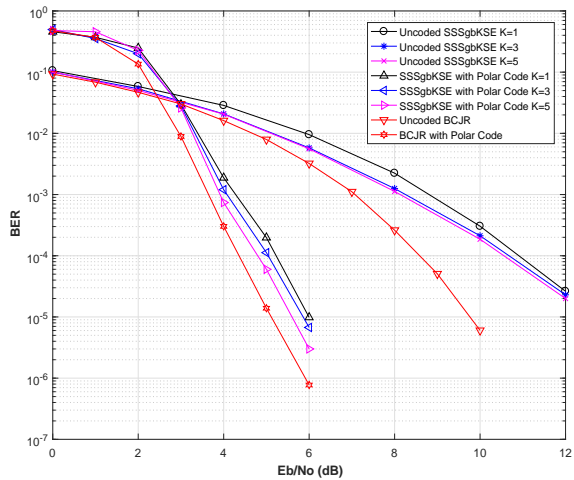}
\caption{Effect of $K\:$ on BER performance at $\tau=$\:0.8. \label{E}}
\end{figure}
Therefore, the polar code is able to reduce performance loss introduced by SSSgb$K$SE detection system by $3.1$\:dB and $1$\:dB when $\tau=$\:0.7 and $\tau=$\:0.6, respectively. 
\par
One can see that polar codes are able to offer additional performance gains to the reduced complexity FTN detection systems. The highest performance gain is provided by polar code is $3.1$\:dB when $\tau=$\:0.7. At  $\tau=$\:0.6 the coding gain is reduced as the SSSgb$K$SE algorithm is not able to provide good LLR values at high ISI condition.

Fig. \ref{E} compares uncoded and coded BCJR performance with those of the  SSSgb$K$SE algorithm for different values of go-back-$K$ values at $\tau=0.8$. For the uncoded case, as the go-back-$K$ value is increased from 1 to 5, the performance of SSSgb$K$SE algorithm slightly improves while the performance loss compared to uncoded BCJR algorithm is aproximately $2$\:dB. However, when polar codes are employed the performance gap between BCJR and SSSgb$K$SE algorithm is quickly reduced to $0.2$\:dB. These results suggest that under light ISI conditions when a channel code is employed there is no need to use complicated FTN detection algorithms as a similar performance can be achieved using a reduced complexity SSSgb$K$SE detection algorithm.
\section{Conclusion}
FTN signaling is a promising signaling method which provides high spectral efficiency to meet the demands of next generation communication systems. In this paper, we investigated the performance of low complexity FTN detection algorithm, i.e., SSSgb$K$SE, when using polar codes. The BER performance of low complexity detection algorithm SSSgb$K$SE is compared against the BCJR detection method under different ISI scenarios and go-back-$K$ values for the coded and uncoded FTN systems. The simulation results reveal that polar coded SSSgb$K$SE detection system achieves higher coding gains compared to polar coded BCJR detection for light ISI scenarios. Consequently, polar codes make up the performance losses introduced by low complexity SSSgb$K$SE algorithm and help to bridge the performance gap with the BCJR algorithm.
\section{Acknowledgements}
H. Yanikomeroglu would like to thank Nokia Bell Labs for their support.

\end{document}